\documentclass[12pt]{article}

\usepackage{scicite}

\usepackage{times}

\usepackage{stackrel}

\usepackage{physics}

\usepackage{amsfonts,amsmath,amssymb,bm,bbm,mathtools,siunitx,mathrsfs}

\usepackage[cal=boondoxo]{mathalfa}

\usepackage{adjustbox}

\DeclareMathAlphabet{\mathpzc}{OT1}{pzc}{m}{it}
\DeclareMathAlphabet\mathbfcal{OMS}{cmsy}{b}{n}

\newenvironment{sciabstract}{%
\begin{quote} \bf}
{\end{quote}}

\title{Phonon Induced Spin Dephasing Time of Nitrogen Vacancy Centers in Diamond from First Principles}

\author
{Jacopo Simoni$^{1\ast}$, Vsevolod Ivanov$^{1,2}$, Thomas Schenkel$^{2}$, Liang Z. Tan$^{1}$\\
\\
\normalsize{$^{1}$Molecular Foundry, Lawrence Berkeley National Laboratory, Berkeley, CA 94720, USA.}\\
\normalsize{$^{2}$Accelerator Technology and Applied Physics Division,Lawrence Berkeley National Laboratory,}\\
\normalsize{Berkeley, CA 94720, USA.}\\
\\
\normalsize{$^\ast$To whom correspondence should be addressed; E-mail: jsimoni@lbl.gov}
}

\date{}

\begin{document} 


\baselineskip24pt


\maketitle


\begin{sciabstract}
    Spin qubits with long dephasing times are an essential requirement for the development of new quantum technologies and have many potential applications ranging from quantum information processing to quantum memories and quantum networking. 
    Here we report a theoretical study and the calculation of the spin dephasing time of defect color centers for the negatively charged nitrogen vacancy center in diamond. We employ ab initio density functional theory to compute the electronic structure, and extract the dephasing time using a cumulant expansion approach. 
    We find that phonon-induced dephasing is a limiting factor for $T_2$ at low temperatures, in agreement with recent experiments that use dynamical decoupling techniques.
    This approach can be generalized to other spin defects in semiconductors, molecular systems, and other band gapped materials.
\end{sciabstract}

{\bf Teaser: } We present calculations of dephasing times in nitrogen vacancy centers

% In setting up this template for *Science* papers, we've used both
% the \section* command and the \paragraph* command for topical
% divisions.  Which you use will of course depend on the type of paper
% you're writing.  Review Articles tend to have displayed headings, for
% which \section* is more appropriate; Research Articles, when they have
% formal topical divisions at all, tend to signal them with bold text
% that runs into the paragraph, for which \paragraph* is the right
% choice.  Either way, use the asterisk (*) modifier, as shown, to

\section{Introduction}
The coherent manipulation of quantum mechanical systems is of fundamental importance in science and engineering\cite{Chatterjee_2020,Kjaergaard_2020}. Solid state quantum systems, in particular, constitute an interesting and very prominent realization of qubits for quantum information processing\cite{Weber_2010,Zhang_2020,Wolfowicz_2021}. The main issue that needs to be overcome is that in contrast to isolated quantum systems, solid state systems tend to have strong coupling with environmental degrees of freedom with consequent loss of coherence of the quantum mechanical state.\\
%
%\myFig{0.6}{0.6}{true}{0}{NV-center.png}{Structure of the NV defect, with the gradients of the $zz$ zero field splitting component, ${\grad}_{aj}{\mathcal{D}}_{zz}$, represented by arrows. The vector is much larger in magnitude for the three carbons close to the vacancy.}{Fig:NVc}
%
Color centers are fluorescent lattice defects made of one or several impurity atoms or vacant atomic sites embedded in the crystal lattice. A particularly prominent example is given by the Nitrogen-Vacancy (NV) defect center in diamond\cite{Schirhagl_2014,Doherty_2013,Lenef_1996,Larsson_2008,Sangtawesin_2014}, that has attracted a lot of attention in the recent years thanks to its long coherence time, optical initialization, and read out\cite{Mizuochi_2009,Choi_2012}. %

The dephasing rate of NV center samples is often controlled by nuclear spin impurities, resulting in coherence times in the range of ms - $\mu$s\cite{Bauch_2018,Bauch_2020,Lin_2021}. However, dynamical decoupling techniques extend coherence time to the order of seconds, where the limiting factor to coherence has been suggested to be spin-phonon interactions instead of nuclear spin impurities~\cite{BarGill_2013}. While the theory of dephasing from nuclear spin impurities has been recently developed~\cite{Seo_2016}, a complete first-principles theory for spin-phonon dephasing is lacking for color centers. With improvements in dynamical decoupling techniques, sample fabrication and purification, and the possibility of room temperature operation, it is desirable to develop a predictive understanding of phonon-induced dephasing. 
Additionally, indications that phonon-induced dephasing introduce non-Markovianity into the system~\cite{Norambuena_2020} necessitate further study of phonon-induced processes in these systems.

In this work we demonstrate via first-principles calculations, that phonon-induced dephasing is a limiting factor for $T_2$ measured by Hahn echo with dynamical decoupling at low temperatures. This additional contribution arises from phonon-induced fluctuations of color center energy levels.  
In general, the total decoherence rate $\Gamma = 1/T_2$ can be written as a sum of four main contributions
%
%\begin{equation}\label{Eq:T2}
% \frac{1}{T_2} = \frac{1}{2T_1} + \frac{1}{T_2^*\textsf{[pure]}} + %\frac{1}{T_2^*\textsf{[disorder]}} +
% \frac{1}{T_2^*\textsf{[QQ]}}\, ,
%\end{equation}
\begin{equation}\label{Eq:T2}
 \Gamma = \frac{1}{T_2} = \frac{1}{2T_1} + \Gamma\textsf{[pure]} + \Gamma\textsf{[disorder]} +
 \Gamma\textsf{[QQ]}\, ,
\end{equation}
where $T_1$ indicates the population relaxation time and defines the limiting decoherence time due to energy relaxation\cite{Norambuena_2018,Jarmola_2012,Gugler_2018,Lunghi_2022} that results from the mixing of the energy levels. 
%$T_2^*$ includes all the remaining contributions to decoherence like pure dephasing and ensemble dephasing. 
The decoherence rate $\Gamma$ refers to the total coherence of the superposition state $(\ket{0}+\ket{1})/\sqrt{2}$ and depends on the specific experimental setup used, as some of the remaining contributions can be removed with an appropriate measurement sequence\cite{Zopes_2017}. 
The terms aside from $T_1$ in Eq.~(\ref{Eq:T2}) represent all other contributions to the decoherence without net energy exchange between the qubit and the environment. As such, $T_2$ in general demands for a harder theoretical treatment compared to $T_1$. \\
Whereas the $T_1$ time for NV centers at cryogenic temperatures can be as long as $8\,$hr\cite{Astner_2018}, $T_2$ was found to be limited, at higher temperatures, to approximately $0.5T_1$\cite{BarGill_2013} and the reported longest measured $T_2$ at low temperature is of the order of $1$ second\cite{Abobeih_2018}. These results indicate that a better understanding of decoherence in solid state qubits requires theoretical methods to directly access the other contributions beyond $T_1$ given the importance of accounting for all the different sources of dephasing. The knowledge of $T_1$ alone is not sufficient to estimate $T_2$.\\
Several dephasing mechanisms are known to contribute beyond population relaxation\cite{Bauch_2018}, here, in order to simplify the problem we distinguish three main groups of dephasing processes and assume that they contribute additively to the total dephasing (Eq.~\ref{Eq:T2}).
%
%\begin{equation}
%    \frac{1}{T_2^*} = \frac{1}{T_2^*\textsf{[pure]}} + %\frac{1}{T_2^*\textsf{[ensemble]}}\,,
%\end{equation}
%
$\Gamma\textsf{[pure]}$ is the pure dephasing rate coming exclusively from unavoidable interactions of a single color center with the environment under the assumption of complete homogeneity. This is the case of a single qubit defect in a perfect crystal lattice with no imperfections and dephasing due only to interactions with atomic vibrations and the radiation field. $\Gamma\textsf{[QQ]}$ is instead the dephasing of the quantum system due to the interaction with other qubits in its environment~\cite{Acosta_2009}. This second term can be reduced greatly by isolating the $NV^{-}$ center from other nitrogen defects in diamond. Finally, each $NV^{-}$ in the ensemble will have a slightly different spin excitation frequency due to local spatial magnetic inhomogeneities causing the qubit to lose coherence during the evolution, a typical example is given by spatial inhomogeneities of the local magnetic field caused by nuclear spin impurities. This term is represented by the $\Gamma\textsf{[disorder]}$ contribution.\\ 
NV centers are often employed for high precision magnetometry and sensing\cite{Degen_2008,Taylor_2008,Abeywardana_2014}, which is usually achieved through the excitation of the spin state evolved under the application of an external magnetic field. Here we consider two quite common methodologies, the Ramsey sequence\cite{vanOort_1990,Epstein_2005} and the Hahn-echo measurements\cite{Carr_1954,Meiboom_1958,Childress_2006}. In the Ramsey sequence of pulses, the qubit is initialized to the superposition state $(\ket{0}+\ket{1})/\sqrt{2}$ and then rotated through the application of an external magnetic field $B$ for a time interval $\tau$. The quantum superposition state at time $\tau$ is then written as $(\ket{0}+e^{i\phi}\ket{1})/\sqrt{2}$ and $\phi$ is the phase acquired by the spin qubit. Finally a $\pi/2$ microwave pulse is applied, projecting the state back to the quantization axis. The Hahn echo sequence instead applies a microwave $\pi$ pulse during the magnetic field evolution phase, flipping the spins and causing a refocusing of the signal. Other decoupling schemes apply several $\pi$ pulses with sensible improvement of the coherence $T_2$ time\cite{Ryan_2010,Naydenov_2011,VdSar_2012}.\\
%In \onlinecite{} optically detected magnetic resonance (ODMR) spectroscopy is performed on $NV^-$ centers with a magnetic field applied along the (111) axis of the center to lift the $m_{\rm s}=\pm 1$ degeneracy. A pulse was first applied to initialize the system to the $m_{\rm s}=0$ state, while a second microwave pulse excites the transition between $m_{\rm s}=0$ and $m_{\rm s}=-1$ states.\\
Previous calculations of dephasing times were mostly applied to semiconducting systems\cite{Liu_2013}, quantum dots\cite{Kamisaka_2006,Palato_2020}, and even biological systems\cite{Mallus_2016}. In these works molecular dynamics (MD) simulations are run at different temperatures and the energy gap fluctuations are obtained using density functional theory at specific configurations along the MD trajectories. These results are used to compute the energy fluctuation auto-correlation function that is related to the dephasing function $D(t)$ through the cumulant expansion approximation\cite{Prezhdo_1997}, and used to extract the dephasing time.\\

In the rest of the paper we will focus on the evaluation of spin dephasing times in the nitrogen vacancy center using constrained Density Functional Theory\cite{Hohenberg_1964,Kohn_1965,Kaduk_2012} (cDFT) based calculations. These methods are routinely employed now to study the electronic structure of defects embedded in solids\cite{Zhang_2020,Ivady_2018,Ivanov2022}.\\
We use the cumulant expansion approximation to obtain the dephasing function. However, we do not use MD simulations to extract the energy fluctuations and the phonon modes are directly computed from the DFT data.
The calculations are performed by means of the \texttt{VASP} package\cite{VASP,VASP2}, which is used to compute the Zero Field Splitting (ZFS) and Hyperfine Interaction (HFI) coefficients. The paper is organized as follows: in section (\ref{sec:theory}) we outline the theoretical formalism. The calculation procedure and the results for NV centers are given in section (\ref{sec:calc}), and in section (\ref{sec:concl}) we conclude.

\section{Results}
\subsection{Theory and formalism} \label{sec:theory}
\subsubsection{Dephasing function calculation} \label{sec:depcalc}
The dephasing function $D(t)$ describes the decay of off-diagonal density matrix components as a function of time after the qubit is initialized in a pure state. 
Within a second order cumulant appoximation~\cite{Mukamel_1991}, $D(t)$ 
is obtained from the autocorrelation function of the fluctuations in energy level differences
\begin{equation}\label{Eq:corr}
  C(t) = \big<\delta E(t)\cdot\delta E(0)\big>_{\rm T}\,,
\end{equation}
where $<\ldots>_{\rm T}$ indicates a thermal average at the temperature $T$, $\delta E(t)$ is the fluctuation in the energy level differences. The dephasing function $D(t)$ is then obtained from the knowledge of the auto-correlation function decay time $\tau_c$ and from the parameter $\Delta^2=C(t=0)$. The following expression is valid in the case of an exponentially decaying autocorrelation function
%
%\begin{equation}
%g(t) = \int_0^td\tau_1\int_0^{\tau_1}d\tau_2\,C(\tau_2)\,,
%\end{equation}
%
\begin{equation}
    g(t) = \Delta^2\tau_c^2\bigg[e^{-t/\tau_c}+\frac{t}{\tau_c}-1\bigg]\,,
\end{equation}
\begin{equation}
  D(t) = e^{-g(t)}\,.
\end{equation}
The cumulant expansion approximation works well under the assumption of harmonic approximation for the phonons. In the limit of fast modulation ($\Delta\tau_c<<1$) the dephasing function becomes $D(t)\simeq e^{-t/T}$ with $T^{-1}=\Delta^2\tau_c$. In the limit of slow modulation, we instead have $g(t)\simeq\Delta^2 t^2/2$ and the inhomogeneous linewidth is simply $\Delta$. In order to proceed, we need an expression for the energy fluctuation $\delta E(t)$. This can be derived from the knowledge of the spin Hamiltonian of the system.
\subsubsection{The spin Hamiltonian}
The spin Hamiltonian has the following general form
\begin{equation} \label{Eq:spinhamil}
    \hat{H}_{\rm ss} = \hat{\bf S}\cdot \stackrel{\leftrightarrow}{\mathbfcal{D}}\cdot\hat{\bf S} + \sum_{\rm I}{\bf I}({\bf R}_{\rm I};t)\cdot\stackrel{\leftrightarrow}{\mathbfcal{A}}_{\rm hfi}({\bf R}_{\rm I})\cdot\hat{\bf S} + \gamma_e{\bf B}\cdot\hat{\bf S}\,,
\end{equation}
The first term on the right hand side is the zero field splitting contribution to the spin Hamiltonian. $\hat{\bf S}$ is the spin operator of the system. $\stackrel{\leftrightarrow}{\mathbfcal{D}}$ is a symmetric and traceless $3\times 3$ tensor that in the case of negligible spin orbit interaction, as in our case, is entirely due to dipolar magnetic interactions\cite{Biktagirov_2018}
\begin{equation}
    D_{ij} = \frac{\mu_0 g_e^2\mu_{\rm B}^2}{4\pi}\sum_{a<b}\chi_{ab}\mel*{\Psi_{ab}}{\frac{r^2\delta_{ij}-3r_i r_j}{r^5}}{\Psi_{ab}}\,,
\end{equation}
where $g_e$ is the Land\'e factor and $\mu_{\rm B}$ is the Bohr magneton. $\ket{\Psi_{ab}}$ is the Slater determinant of the two-electrons system, that in our case corresponds to the Kohn-Sham Slater determinant obtained from the solution of the DFT set of coupled equations. $r$ is the distance between the two interacting spins.
The second term on the right hand side of Eq.~(\ref{Eq:spinhamil}) is the hyperfine coupling with nuclear spins ${\bf I}({\bf R}_{\rm I})$. It defines an effective time dependent magnetic field given that the nuclear spins will also evolve in time under an externally applied magnetic field ${\bf B}$.\\
The hyperfine coupling term $\stackrel{\leftrightarrow}{\mathbfcal{A}}_{\rm hfi}$ is given by the sum of the Fermi contact contribution and a dipolar term\cite{Frosch_1952}
\begin{equation}
    \mathcal{A}_{\rm hfi}^{ij}({\bf R}_{\rm I}) = \frac{\mu_0 g_e g_{\rm I}\mu_{\rm B}\mu_{\rm J}}{\expval{S_z}}\bigg[\frac{2}{3}\delta_{ij}\rho_S({\bf R}_{\rm I}) + \frac{1}{4\pi}\int d{\bf r}\frac{\rho_S({\bf r}+{\bf R}_{\rm I})}{r^3}\frac{3 r_i r_j-\delta_{ij}r^2}{r^2}\bigg]\,,
\end{equation}
$\rho_S({\bf R}_{\rm I})$ is the spin electron density located around the atom ${\rm I}$. $g_{\rm I}$ and $\mu_{\rm J}$ are the nuclear Land\'e factor and the nuclear magneton. The last term in Eq.~(\ref{Eq:spinhamil}) is the Zeeman coupling term with $\gamma_e$ electron gyromagnetic ratio.
\subsubsection{Hyperfine and zero field splitting energy fluctuations}
Our formalism requires the evaluation of fluctuations in the spin levels $\delta E(t)$ at time $t$ from the different sources of dephasing. We assume that the qubit is evolved from some linear combination of states $\ket{0}$ and $\ket{1}$ of the spin triplet, implying that $\delta E(t)=\mel*{1}{\delta\hat{H}_{\rm ss}(t)}{1}-\mel*{0}{\delta\hat{H}_{\rm ss}(t)}{0}$ is the fluctuation in the energy difference between the two eigenstates of the spin Hamiltonian. The fluctuation can then be expressed as
\begin{equation} \label{Eq:enfluct}
    \delta\hat{H}_{\rm ss} = \delta\hat{H}_{\rm ss}^{\rm sp-ph} + \delta\hat{H}_{\rm ss}^{\rm sp-nu-ph} + \delta\hat{H}_{\rm ss}^{\rm sp-nu}\,,
\end{equation}
\begin{equation}
    \delta\hat{H}_{\rm ss}^{\rm sp-ph} = \sum_\lambda \sum_{\bf q} \sum_{j;a} u_{\lambda,{\bf q}}(ja;t) \hat{\bf S}\cdot\nabla_{ja}\stackrel{\leftrightarrow}{\mathbfcal{D}}\cdot\hat{\bf S}\,,\nonumber
\end{equation}
\begin{equation}
     \delta\hat{H}_{\rm ss}^{\rm sp-nu-ph} = \sum_{\lambda,{\bf q}} \sum_{j;a} u_{\lambda,{\bf q}}(ja;t) \sum_{\rm I}{\bf I}_0({\bf R}_{\rm I})\cdot\nabla_{ja}\stackrel{\leftrightarrow}{\mathbfcal{A}}_{\rm hfi}({\bf R}_{\rm I})\cdot\hat{\bf S}\,,\nonumber
\end{equation}
\begin{equation}
     \delta\hat{H}_{\rm ss}^{\rm sp-nu} = \sum_{\rm I}\delta{\bf I}({\bf R}_{\rm I};t)\cdot\stackrel{\leftrightarrow}{\mathbfcal{A}}_{\rm hfi}({\bf R}_{\rm I})\cdot\hat{\bf S}\,,\nonumber
\end{equation}
where $u_{\lambda,{\bf q}}(ja;t)$ is the atomic vibration associated with the mode $(\lambda,{\bf q})$ along atom $a$ and direction $j$. $\delta{\bf I}({\bf R}_{\rm I};t)$ is the temporal variation of the nuclear spin as a result of the precessional motion with respect to ${\bf I}_0({\bf R}_{\rm I})$ due to the external magnetic field.
In Eq.~(\ref{Eq:enfluct}) we have defined three main contributions to the fluctuations in the energy levels. 
%The first two involve coupling with phonon modes while the third is due to the interaction between nuclear and electronic spins. The two are spin-phonon contributions are caused by fluctuations in the spin levels due to the underlying atomic vibrations of the crystal. 
The first term describes the phonon-induced fluctuations of the spin-spin coupling tensor $\stackrel{\leftrightarrow}{\mathbfcal{D}}$, and is the most important contribution to $\Gamma\textsf{[pure]}$.
The second term involves the phonon-induced fluctuations of the hyperfine coupling while the last term accounts for the precession of nuclear spins.
Due to their dependence on nuclear spin impurities, the second and last terms ($\delta\hat{H}_{\rm ss}^{\rm sp-nu-ph}$ and $\delta\hat{H}_{\rm ss}^{\rm sp-nu}$) constitute the main contributions to $\Gamma\textsf{[disorder]}$.
From now on we will assume that our system forms a matrix of single $NV^-$ defects located far enough apart that they do not interact with each other. This assumption is valid in the case of low nitrogen concentration. 
This means that we can neglect $\Gamma\textsf{[QQ]}$ and consider how the three terms in Eq.~(\ref{Eq:enfluct}) contribute to $\Gamma\textsf{[pure]}$  and $\Gamma\textsf{[disorder]}$. Their relative importance
%This means that our $T_2$ corresponds essentially to the $T_2^\textsf{[pure]}$ contribution with an additional source of inhomogeneity given by the concentration of $C^{13}$ isotopes. The way the three terms contribute to $T_2^*$ 
changes with the particular experimental setup used, and necessitates individual examination in order to predict the experimentally observed $T_2$ values.
In Tab.~(1), we summarize the results of our calculations, which will be discussed in the following sections. 
\subsection{Nitrogen vacancy center calculations} \label{sec:calc}

The negatively charged NV center is a paramagnetic ground state defect with quantization axis directed along the nitrogen-vacancy axis. The $NV^-$ center is characterized by a spin triplet ground state, $^3A_2$, and a spin triplet excited state $^3E$, with zero field splittings given respectively by $D=2.87\,{\rm GHz}$ and $D=1.42\,{\rm GHz}$, as well as two singlet states, $^1A_1$ and $^1E$\cite{Doherty_2013,Lenef_1996,Larsson_2008}. Upon optical excitation, the NV center shows strong fluorescence, the intensity of which is spin dependent due to spin dependent relaxation via singlet states\cite{Choi_2012}. All the details of the calculations are given in section (\ref{sec:DFT}) while in the next sections we discuss our results in case of a dynamical decoupling sequence and of a Ramsey sequence. From now on we will assume that the system is initialized in its spin triplet ground state configuration.

\subsubsection{Hahn echo dephasing and dynamical decoupling}
%
%\myFig{1}{1}{true}{0}{T2_NVc.pdf}{Evaluation of the pure dephasing time $1/\Gamma\textsf{[pure]}$ due to the first order spin-phonon coupling in absence of nuclear spin impurities. Our result (black line with blue bar indicating the confidence interval) is compared with experimental data from Ref. (A) \onlinecite{BarGill_2013} and Ref. (B) \onlinecite{Abobeih_2018}.}{Fig:03}
%
A fundamental property of the Hahn echo sequence is the removal of inhomogeneous broadening from static or slowly varying magnetic fields. In general, different points in the $NV^-$ matrix will be characterized by different local values of the magnetic field, due to different nuclear spin distributions. This causes strong spatial dephasing that can be inhibited by means of the Hahn echo sequence. The result can be systematically improved by means of more complex decoupling techniques, making the coherence time still longer\cite{BarGill_2013}. However, for AC magnetic fields, these methods are ineffective for improving the dephasing time. 
In (Fig. 2) we show the inverse of the $\Gamma\textsf{[pure]}$ values computed for the $NV^-$ center in diamond over a range of temperatures between $T=10\,K$ and $T=500\,K$ with energy fluctuations caused only by $\delta\hat{H}_{\rm ss}^{\rm sp-ph}$ in Eq.~(\ref{Eq:enfluct}). The so obtained dephasing time is close, at least at low temperature, to the $T_2$ values reported in Ref. \cite{BarGill_2013} and \cite{Abobeih_2018}. In both works, the combination of dynamical decoupling techniques and cryogenic cooling lead to a sensible increase in the observed value of $T_2$ up to almost $1\,s$. 
%These values are close or, in the case of the value reported in \onlinecite{Abobeih_2018}, within the confidence interval of our calculation. 
The blue error bars in (Fig. 2) are determined by the choice of the fitting function model for the energy auto-correlation function. In the case of a simple fit of $C(t)$ to an exponential $Ae^{-t/\tau_c}+B$ we obtain the upper limit of the interval, whereas if we use a sinusoidal modulated exponential, $A \sin(\omega t+\phi)e^{-t/\tau_c} + B$, we obtain the lower limit.\\
%
%\myFig{1}{1}{true}{0}{T2-atr-phr.pdf}{atoms and phonon resolved $1/\Gamma\textsf{[pure]}$ times. (Upper panel) atom-resolved $1/\Gamma\textsf{[pure]}$ is shown as a function of distance from the vacancy, expressed as a fraction of the length of the 3$\times$3$\times$3 supercell. (Lower panel) phonon resolved dephasing time as a function of the phonon frequency.}{Fig:04}
%
Although our computed $1/\Gamma\textsf{[pure]}$ compares remarkably well with the experimental $T_2$ from Ref.~\cite{BarGill_2013} and \cite{Abobeih_2018} at low temperatures ($3\,K$ and $70\,K$), at higher temperatures theory and experiments diverge. This can be in part explained by the fact that our temperature dependence is underestimated, as most of the computed $\Gamma\textsf{[pure]}$ is a result of quantum zero-point fluctuations of the phonon bath. We are, in fact, only considering the second order term of the cumulant expansion (Eq.~\ref{Eq:corr}), and first order variations in the phonon displacement (Eq.~\ref{Eq:enfluct}), leaving out the higher order contributions that account for multi-phonon processes, which are only relevant at high temperatures.\\
In our calculations, in fact, the temperature dependence enters the expression for the energy fluctuations only through the phonon amplitude with a dependence of the form $\sqrt{1+2n_{\rm ph}}$. Most importantly, the measured coherence time $T_2$ is not equivalent to $1/\Gamma\textsf{[pure]}$. Our calculations suggest that at low temperatures, thanks to very long spin relaxation times\cite{Astner_2018}, $T_2$ should be dominated by the pure dephasing term. On the other hand, at higher temperatures, $T_1$ dominates over $1/\Gamma\textsf{[pure]}$ due to its stronger temperature dependence.\\
(Fig. 3) evaluates the contribution of the different atoms in the super cell and of the different phonon modes to $1/\Gamma\textsf{[pure]}$. The upper panel in the figure shows each atom's resolved dephasing time as a function of each atoms distance from the vacancy. This quantity is obtained from Eq.~(\ref{Eq:enfluct}) eliminating the sum over the atom's indices $(j;a)$. We observe a clear trend in the figure with the carbon atoms closer to the vacancy having a greater effect on the $\Gamma\textsf{[pure]}$ linewidth compared to the atoms farther away. This phenomenon can be understood qualitatively by considering that the atom-resolved zero field splitting gradient ${\grad}_{aj}{\stackrel{\leftrightarrow}{\mathbfcal{D}}}$ is much higher in magnitude for a few carbon atoms located around the vacancy. In (Fig. 1) we can clearly distinguish three carbon atoms corresponding to the three red points of (Fig. 3) with a much higher magnitude of the $\grad_{a}\stackrel{\leftrightarrow}{\mathbfcal{D}}$ vectors compared to other atoms (see also (Fig. SM1)). The three vectors are oriented toward the vacancy with the same trigonal symmetry posessed by the defect. This is in agreement with the observation that $T_1$ times in NV centers are mostly determined by the local vibrational properties around the defect center\cite{Astner_2018}. We predict that the first shell (red points in (Fig. 3)) and a few carbons in the second shell around the vacancy have equally strong contribution to the overall dephasing, approximately an order of magnitude stronger than the other atoms. The nitrogen atom (blue point in figure), despite being a nearest-neighbor of the vacancy, does not contribute as much because of its comparatively minor electron spin density.\\
The lower panel of the figure shows instead the contribution to $\Gamma\textsf{[pure]}$ coming from the different phonon modes. Each point in the figure corresponds to a vibrational mode; orange colored points are more localized close to the vacancy compared to darker points. We distinguish two frequency bands which contribute strongly to $\Gamma\textsf{[pure]}$. The broad band at 10 - 20 THz is the contribution of local-continuum resonances, while the sharp band at 40 THz comes from local modes~\cite{Alkauskas_2014}. In general, the contribution to $\Gamma\textsf{[pure]}$ increases with the local character of the mode. \\
%The low frequency acoustic modes are more localized around the defect atoms, since there is a higher concentration of yellow points, compared to the high frequency modes. These modes also have shorter dephasing times and are the ones mostly responsible for shortening the value of $T_2^*\textsf{[pure]}$.
%
%\myFig{1}{1}{true}{0}{T2_NVc+HFI.pdf}{comparison between $1/\Gamma$ due to the spin-phonon term and the HFI-spin-phonon term in seconds at different $C^{13}$ concentrations.}{Fig:04b}
%
In (Fig. 4) we finally consider the effect of the hyperfine coupling. We separate the contribution of $\delta\hat{H}_{\rm ss}^{\rm sp-nu-ph}$ from that of $\delta\hat{H}_{\rm ss}^{\rm sp-ph}$, as in Eq.~(\ref{Eq:enfluct}). Due to the fast vibrational dynamics the effect of $\delta\hat{H}_{\rm ss}^{\rm sp-nu-ph}$ is not mitigated by dynamical decoupling techniques and it disappears only in the limit of very low $C^{13}$ concentrations. In our simulations we apply an external static magnetic field ${\bf B}$ along the spin quantization axis, which defines a preferential alignment axis for the nuclear spins, and then average over $32$ possible spin configurations, enough to converge on the final $1/\Gamma$ value. In each configuration the nuclear spins are associated to a random set of atoms in the simulation box. At finite temperatures the nuclear spin direction has finite probability of not being aligned with the applied magnetic field. The direction of the nuclear spins, in the different configurations, are selected randomly from a Gaussian distribution centered on the {\bf B} field direction and with a width proportional to the temperature of the system. This has the effect of making the $1/\Gamma$ time sightly longer compared to the zero temperature value. However, the most important contribution to $1/\Gamma$ comes from the concentration of $C^{13}$ isotopes. At low concentrations $1/\Gamma[{\rm disorder}]\simeq 10^8\,s$, while at higher concentrations $1/\Gamma[{\rm disorder}]$ decreases by a few orders of magnitude, which is not sufficient to make this effect observable compared to the energy fluctuations due to the simple spin-phonon term $\delta\hat{H}_{\rm ss}^{\rm sp-ph}$.

\subsubsection{Ramsey sequence dephasing times}
%
%\myFig{1}{1}{true}{0}{T2_star.pdf}{Calculation of the inhomogeneous dephasing $1/\Gamma[\textsf{disorder}]$ in seconds obtained at different magnetic field amplitudes $50\,G$ (circle points), $100\,G$ (squares), $500\,G$ (stars) and $1000\,G$ (triangles). The two additional data points are experimental data from Ref. (C) \onlinecite{Mizuochi_2009}}{Fig:05}
%
The Ramsey sequence coherence times  (commonly referred to as $T_2^*$) are fundamentally determined by the strength of the hyperfine interaction close to the defect center. It depends on two extrinsic parameters, the applied magnetic field strength and the concentration of nuclear spin impurities. This is an overestimate of the experimental $T_2$ since other effects could be at play that are not considered here like the magnetic interaction between different defects\cite{Bauch_2020},temperature fluctuations and strains.\\ 
In (Fig. 5) we compute the $1/\Gamma\textsf{[disorder]}$ time as a function of the concentration of carbon magnetic impurities for different applied magnetic field amplitudes. The calculations are performed in all the different cases by randomly selecting $128$ nuclear spin configurations; the nuclear spins in each configuration are evolved under the effect of the externally applied magnetic field and of the electronic spins coupled through the hyperfine tensor. The spin fluctuations are then computed according to Eq.~(\ref{Eq:enfluct}) for each configuration by isolating the $\delta\hat{H}_{\rm ss}^{\rm sp-nu}$ term. The $1/\Gamma\textsf{[disorder]}$ time of the ensemble is obtained by taking the average of the energy fluctuation functions $\delta E$ from the different configurations. To understand the distribution within the ensemble, we also compute $\Gamma\textsf{[disorder]}$ for each configuration alone and plot an error bar depicting the standard deviation of the distribution. The standard deviation is bigger at low concentrations due the low number of nuclear spins contributing to the dynamics. At low concentrations we find the longest dephasing times, approaching $0.1\,ms$, while at high concentrations we converge to values of $1/\Gamma\textsf{[disorder]}$ below $1\,\mu s$. The application of stronger magnetic fields to the same random configuration of nuclear spins lowers the Ramsey sequence dephasing times due to the higher spin precession frequencies, as seen in (Fig. 5). On the other hand, applied magnetic fields also tend to align nuclear spins, which would counteract this effect. This has been discussed elsewhere~\cite{Hall_2014, Seo_2016} and it is not considered here since we assume that our starting spin configuration is randomly distributed and not aligned to the applied field.

\section{Discussion} \label{sec:concl}
We have computed the pure dephasing time (at the second order cumulant approximation) and part of the ensemble dephasing coming from $C^{13}$ isotopic impurities in diamond $NV^-$ centers. These results indicate the importance of accounting for various dephasing mechanisms in the calculation of the full decoherence time in solid state qubits. The application of dynamical decoupling techniques can improve the decoherence time but we find that at low temperatures the spin-phonon contribution to the pure dephasing time ($1/\Gamma\textsf{[pure]}$) sets an upper limit for the decoherence time in agreement with recent experiments. At higher temperatures spin-phonon relaxation becomes the dominant contribution to the decoherence due to the weak temperature dependence of the spin-phonon pure dephasing term. The disorder induced dephasing time is the dominant contribution at high impurity concentrations, while at low concentrations it sets a limit of the order of few milliseconds\cite{BarGill_2013} that can be overcome by means of dynamical decoupling techniques. These results suggest that phonon-induced dephasing should be evaluated in the design of new color centers as a potential limiting factor to coherence in particular at low temperatures.

\section{Materials and Methods} \label{sec:DFT}

The electronic structure of the negatively charged $NV^-$ center is computed using \texttt{VASP}\cite{VASP,VASP2} and PBE functionals\cite{Perdew_1996}. The simulations are performed at the $\Gamma$ point using a $3\times 3\times 3$ super cell with a total of $215$ atoms (Figure 1).%\ref{fig:NNc}
An increase of the number of {\bf k} points does not produce any significant change in the zero field splitting or hyperfine tensor values. The ground state is a spin triplet with a zero field splitting $D = 2.97\,$GHz that is in good agreement with previous calculations and the experimentally reported value of $2.87\,$GHz\cite{Sangtawesin_2014,Childress_2006,VdSar_2012}.\\
The vibrational modes of the system are then computed using the \texttt{phonopy} package\cite{Togo_2015} and used into Eq.~(\ref{Eq:enfluct}) to extract the energy fluctuations. A $8\times 8\times 8$ {\bf q}-vectors grid with $244$ irreducible {\bf q}-points was required to achieve convergence in the summation of the phonon wave vectors. In a supercell with $215$ atoms the number of vibrational modes is $645$. The hyperfine coupling is also computed using \texttt{VASP}. The ZFS and the HFI gradients, $\grad_{aj}\stackrel{\leftrightarrow}{\mathbfcal{D}}$ and $\grad_{aj}\stackrel{\leftrightarrow}{\mathbfcal{A}}_{\rm hfi}$, are obtained by means of a finite difference real space approach where each atom in the simulation box is separately displaced to a new position $R_{0,x}\pm dx$ along each of the three Cartesian directions. The ground state DFT calculation is then repeated for each of these new atomic configurations. The typical displacements employed here are of the order of $dx=10^{-3}$\AA. Once we computed the gradients we evaluate $\delta E(t)$ and its auto-correlation function. We can then extract the different $\Gamma$ contributions following the procedure outlined in section (\ref{sec:depcalc}). The calculation of the energy fluctuations in the spin-nuclear term does not require the knowledge of the phonon modes and it is less computationally demanding.

% Your references go at the end of the main text, and before the
% figures.  For this document we've used BibTeX, the .bib file
% scibib.bib, and the .bst file Science.bst.  The package scicite.sty
% was included to format the reference numbers according to *Science*
% style.

%BibTeX users: After compilation, comment out the following two lines and paste in
% the generated .bbl file. 

\bibliography{scibib}

\bibliographystyle{Science}

\section*{Acknowledgments}
\subsection*{Funding}
This work was supported by the Office of Science, Office of Fusion Energy Sciences, of the U.S. Department of Energy, under Contract No. DE-AC02-05CH11231.  JS, VI and LZT were also supported by the Molecular Foundry, a DOE Office of Science User Facility supported by the Office of Science of the U.S. Department of Energy under Contract No. DE-AC02-05CH11231.  This research used resources of the National Energy Research Scientific Computing Center, a DOE Office of Science User Facility supported by the Office of Science of the U.S. Department of Energy under Contract No. DE-AC02-05CH11231.

\subsection*{Author contributions}
JS and LZT worked on the theory and methodologies, JS and VI worked on the calculations and data visualization, LZT and TS supervised the work and JS, VI, LZT, TS wrote the manuscript. 

\subsection*{Competing interests}
The authors declare no competing financial interest.

\subsection*{Data and materials availability}
The data supporting the different figures and tables are available from the corresponding author upon reasonable request.

\clearpage

\section{Figures and Tables}

% For your review copy (i.e., the file you initially send in for
% evaluation), you can use the {figure} environment and the
% \includegraphics command to stream your figures into the text, placing
% all figures at the end.  For the final, revised manuscript for
% acceptance and production, however, PostScript or other graphics
% should not be streamed into your compliled file.  Instead, set
% captions as simple paragraphs (with a \noindent tag), setting them
% off from the rest of the text with a \clearpage as shown  below, and
% submit figures as separate files according to the Art Department's
% instructions.

\begin{figure}[h]
    \centering
    \includegraphics[width=0.8\textwidth]{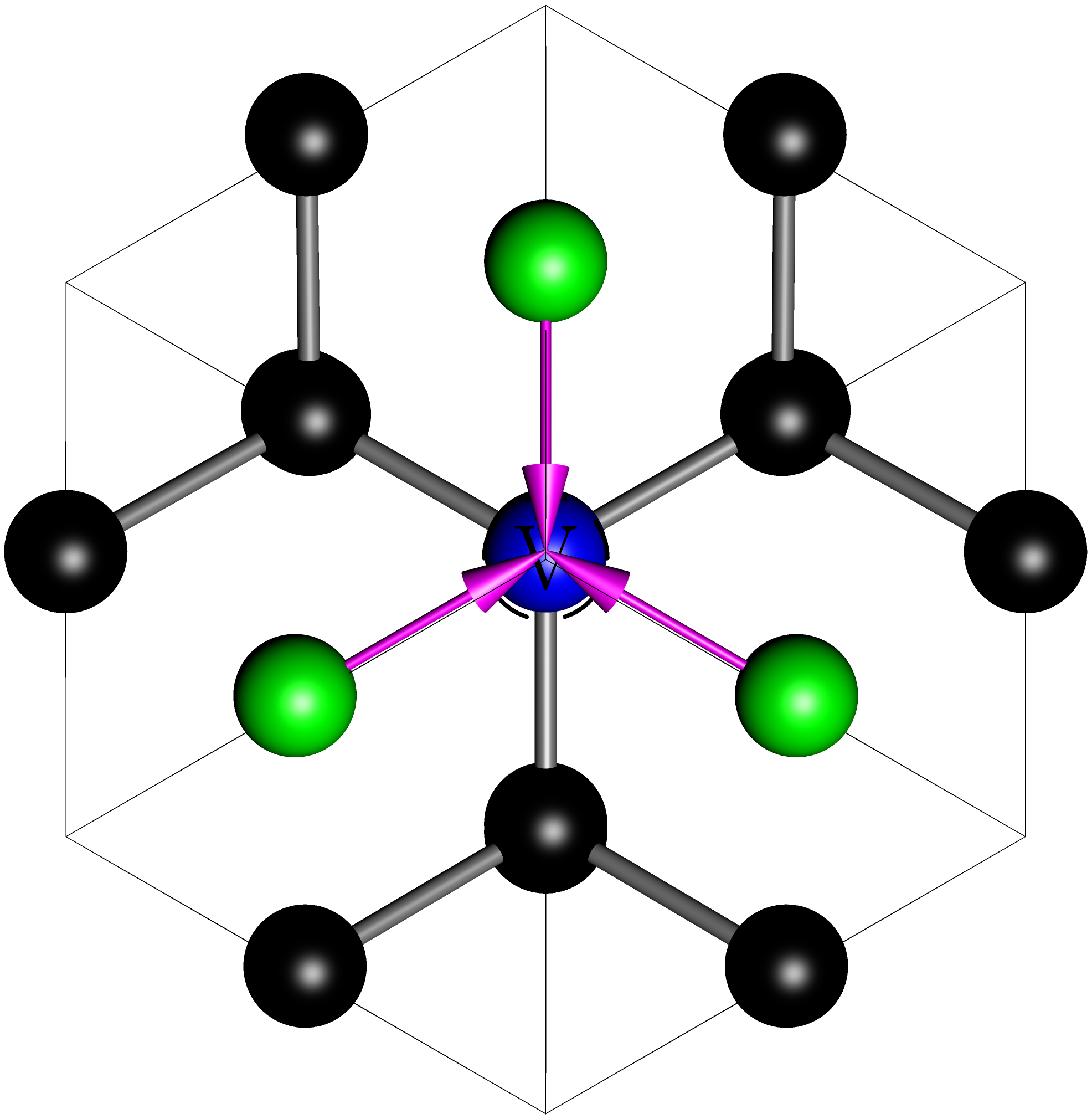}
\end{figure}
\noindent {\bf Fig. 1.} Structure of the NV defect, with the gradients of the $zz$ zero field splitting component, ${\grad}_{aj}{\mathcal{D}}_{zz}$, represented by arrows. The vectors associated to the three carbons surrounding the vacancy have the same trigonal symmetry of the defect.

\clearpage

\begin{figure}[h]
    \centering
    \includegraphics[width=1.0\textwidth]{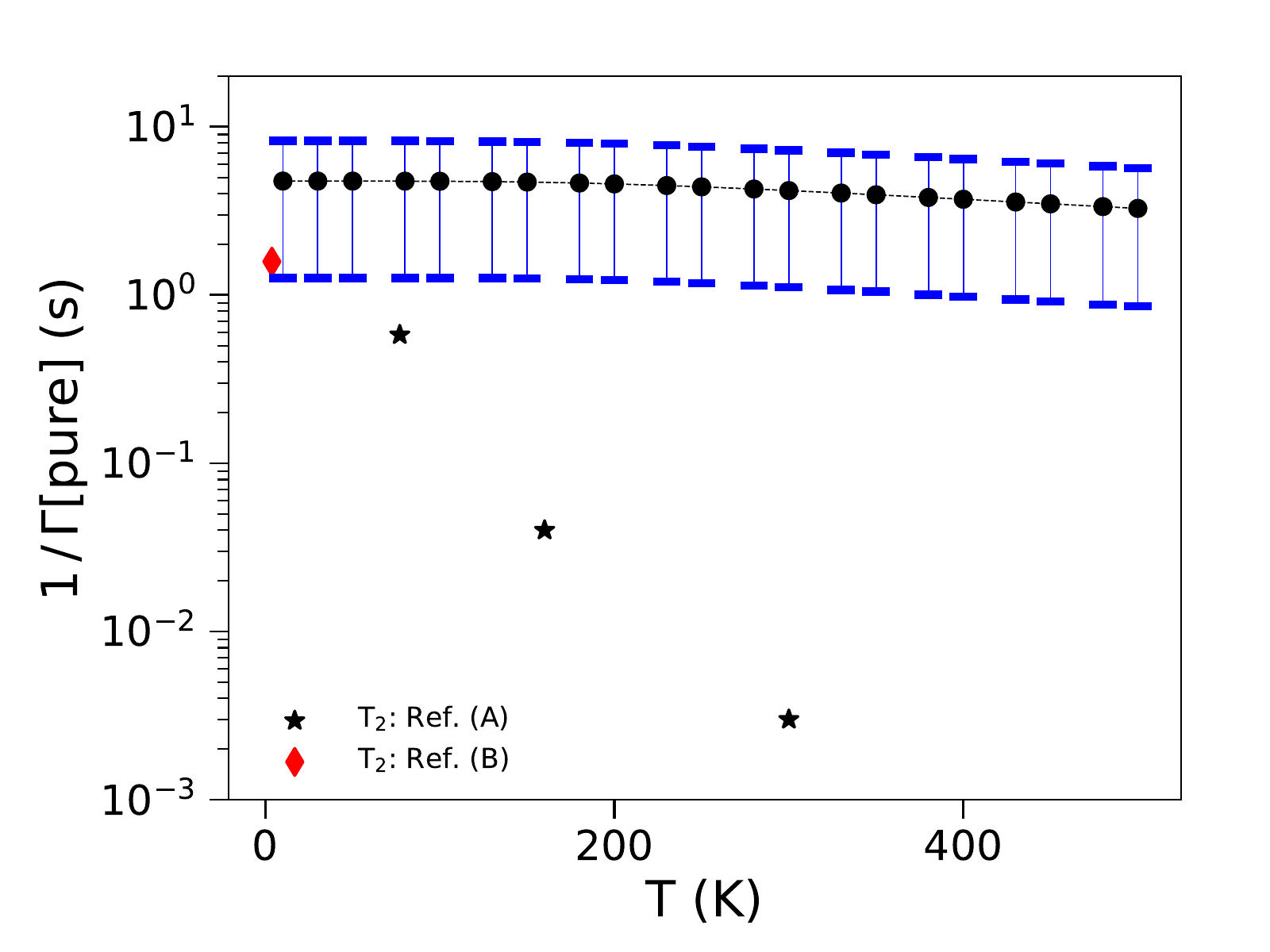}
\end{figure}
\noindent {\bf Fig. 2.} Evaluation of the pure dephasing time $1/\Gamma\textsf{[pure]}$ due to the first order spin-phonon coupling in absence of nuclear spin impurities. Our result (black line with blue bar indicating the confidence interval) is compared with experimental data from Ref. (A) \cite{BarGill_2013} and Ref. (B) \cite{Abobeih_2018}.

\clearpage

\begin{figure}[h]
    \centering
    \includegraphics[width=0.9\textwidth]{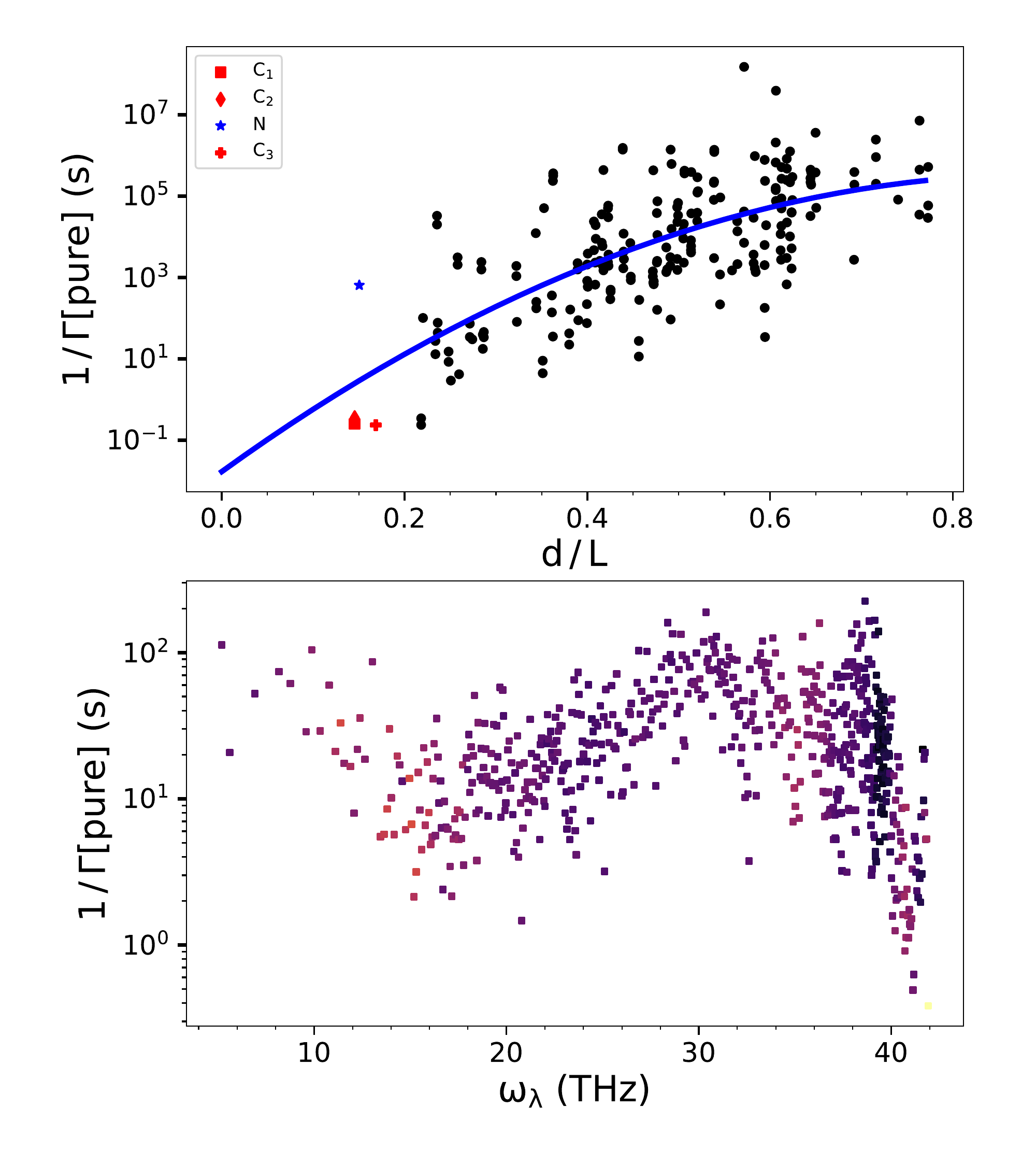}
\end{figure}
\noindent {\bf Fig. 3.} Atoms and phonon resolved $1/\Gamma\textsf{[pure]}$ times. (Upper panel) atom-resolved $1/\Gamma\textsf{[pure]}$ is shown as a function of distance from the vacancy, expressed as a fraction of the length of the 3$\times$3$\times$3 supercell. (Lower panel) phonon resolved dephasing time as a function of the phonon frequency.

\clearpage

\begin{figure}[h]
    \centering
    \includegraphics[width=1.0\textwidth]{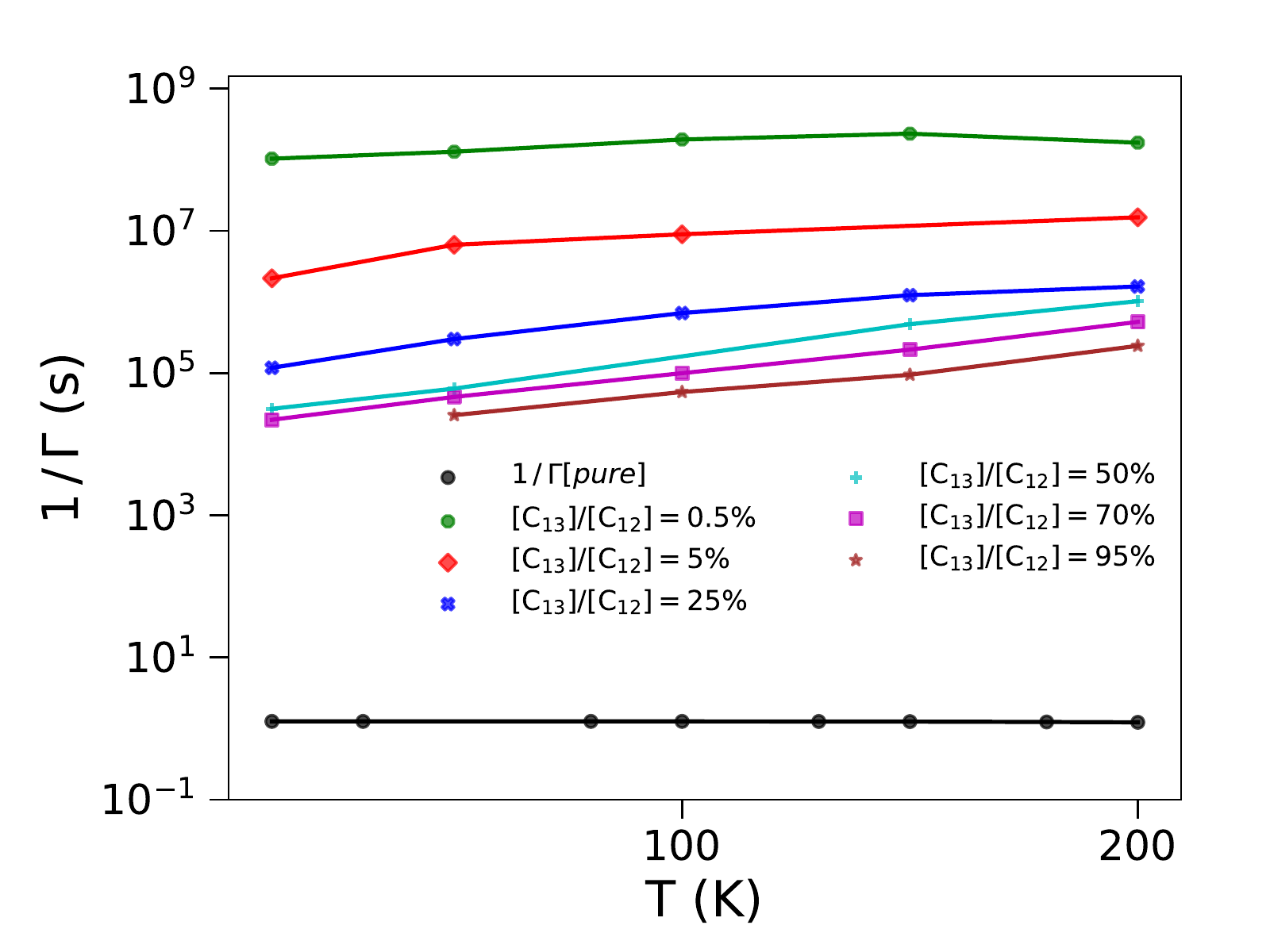}
\end{figure}
\noindent {\bf Fig. 4.} Comparison between $1/\Gamma$ due to the spin-phonon term and the HFI-spin-phonon term in seconds at different $C^{13}$ concentrations.

\clearpage

\begin{figure}[h]
    \centering
    \includegraphics[width=1.0\textwidth]{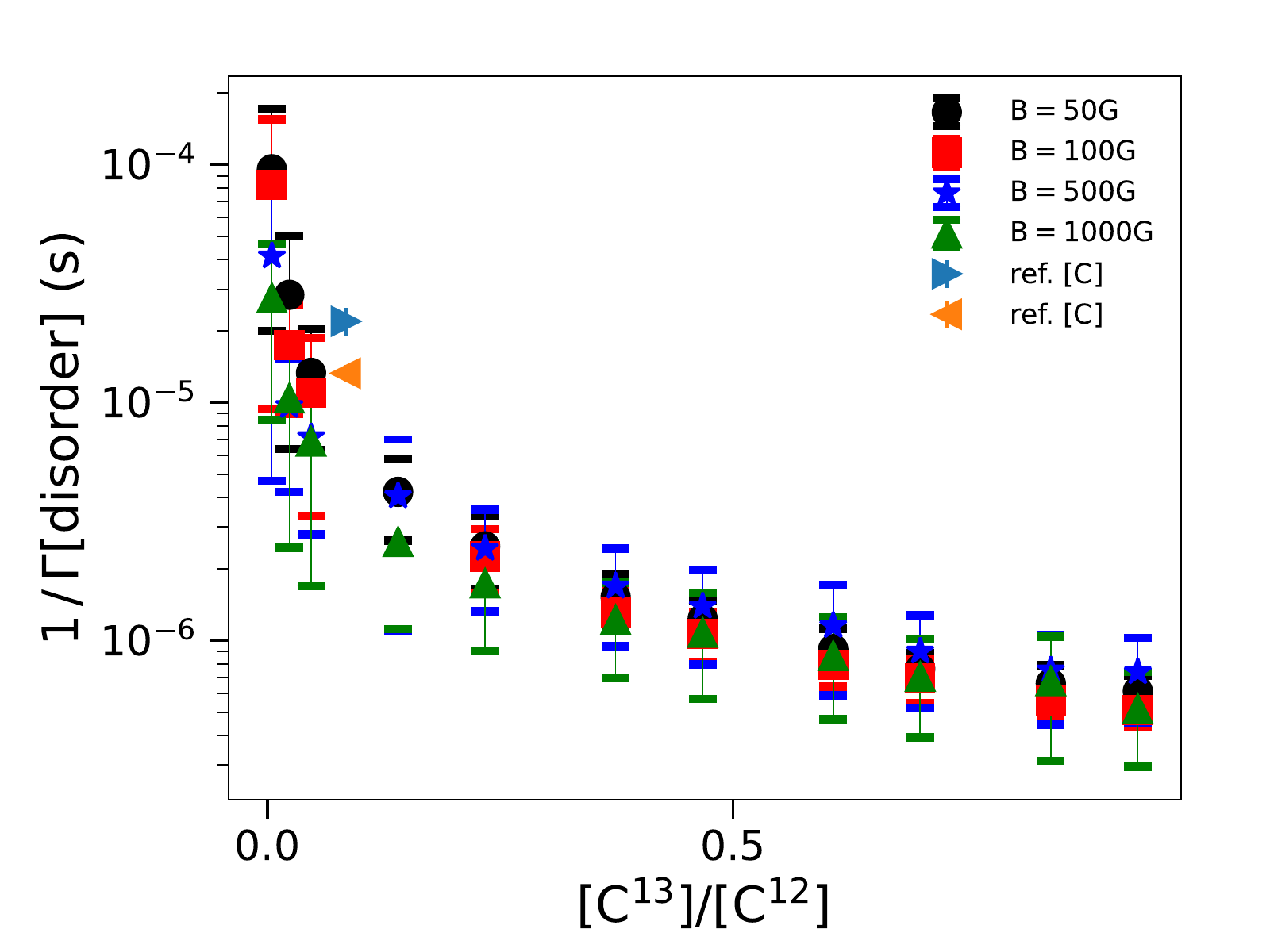}
\end{figure}
\noindent {\bf Fig. 5.} Calculation of the inhomogeneous dephasing $1/\Gamma[\textsf{disorder}]$ in seconds obtained at different magnetic field amplitudes $50\,G$ (circle points), $100\,G$ (squares), $500\,G$ (stars) and $1000\,G$ (triangles). The two additional data points are experimental data from Ref. (C) \cite{Mizuochi_2009}.

\clearpage

\begin{table}[ht]
\centering % used for centering table
\begin{adjustbox}{width=1.0\textwidth}
\begin{tabular}{c c c c} % centered columns (4 columns)
\hline\hline %inserts double horizontal lines
$\Gamma\textsf{[pure]}^{-1}$ & $4.8\pm 3.5$ s & homogeneous & irreversible \\
$\Gamma_A\textsf{[disorder]}^{-1}$ & $10^{8}$ s & inhomogeneous & irreversible \\
$\Gamma_B\textsf{[disorder]}^{-1}$ & $10^{-4}$ s & inhomogeneous & reversible \\
%   & Theory & Experiment \\ [0.5ex] % inserts table
%heading
%\hline % inserts single horizontal line
%NV & 50 & 2.87\cite{Schirhagl_2014,Childress_2006}\\ % inserting body of the table
%GB & 47 & 877\\
%GA & 45 & 300\\ [1ex] % [1ex] adds vertical space
\hline %inserts single line
\end{tabular}
\end{adjustbox}
\end{table}

\noindent {\bf Tab. 1.} Summary of dephasing mechanisms considered in this work, together with computed dephasing times and associated properties. Homogeneous dephasing refers to the fast-modulation case where the decay rate of energy level fluctuations is fast compared to the amplitude of fluctuations. The reversible dephasing can be removed by dynamical decoupling. Label (A) indicates (sp-ph-nu) and (B) (sp-nu). We assume a concentration of nuclear spins of $0.5\%$ and for the (sp-nu) term the applied magnetic field has a magnitude of $50\,G$.

\clearpage

%Here you should list the contents of your Supplementary Materials -- below is an example. 
%You should include a list of Supplementary figures, Tables, and any references that appear only in the SM. 
%Note that the reference numbering continues from the main text to the SM.
% In the example below, Refs. 4-10 were cited only in the SM.

\section{Supplementary materials}
%Materials and Methods\\
%Supplementary Text\\
%Figs. S1 to S3\\
%Tables S1 to S4\\
%References \textit{(4-10)}

\begin{figure}[h]
    \centering
    \includegraphics[width=0.8\textwidth]{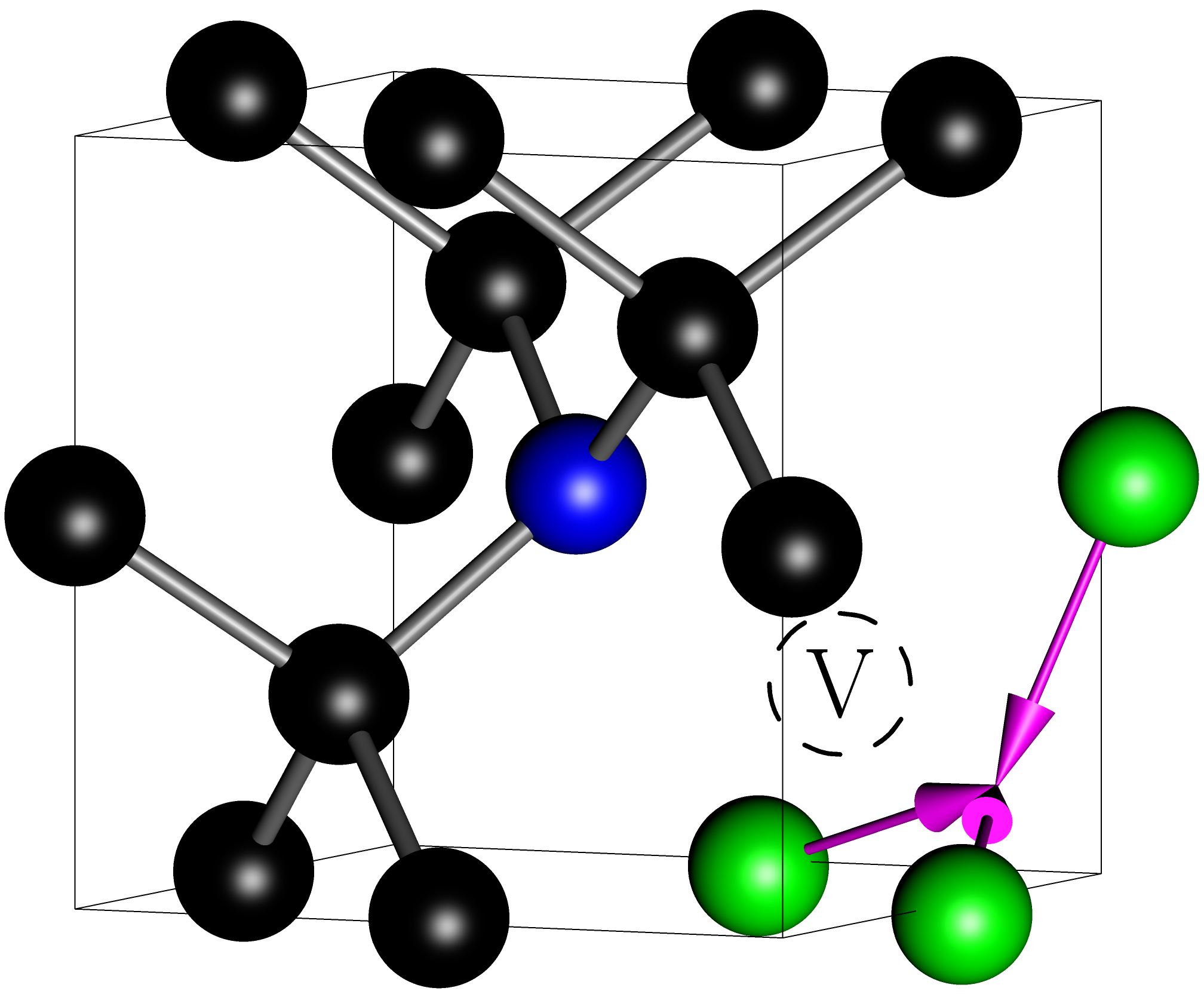}
\end{figure}
\noindent {\bf Fig. SM1.} Structure of the NV defect, with the gradients of the $zz$ zero field splitting component, ${\grad}_{aj}{\mathcal{D}}_{zz}$, represented by arrows. The gradient for the three carbons close to the vacancy is much larger in magnitude compared to the other atoms.

\end{document}